\documentclass[twocolumn,amsmath,amssymb]{revtex4}

\begin{document}

\title{Entanglement Dynamics and Quantum State Transport in Spin chains}
\author{V. Subrahmanyam}
\affiliation{
%\address{
Max-Planck-Institut fuer Physik komplexer Systeme,
Noethnitzer Str.38, Dresden 01187 Germany\\
and 
Department of Physics, Indian Institute of Technology, 
Kanpur 208016, India}
\altaffiliation[Permanent address, E-mail: vmani@iitk.ac.in\\]{}

\date{\today}

\begin{abstract} 
We study the dynamics of a Heisenberg-XY spin chain with an unknown state
coded into one qubit or a pair of
entangled qubits, with the rest of the spins being in a polarized state.
The time evolution involves magnon excitations, and through them the
entanglement is
transported across the channel. For a large number of qubits, explicit
formulae for the concurrences, measures for two-qubit entanglements, and 
the fidelity for recovering the state some distance away are calculated 
as functions of time. Initial states with an entangled pair of qubits show 
better fidelity, which
takes its first maximum value at earlier times, compared to initial states
with no entangled pair. In particular initial states with
a pair of qubits in an unknown state $\alpha\uparrow\uparrow+\beta\downarrow
\downarrow$ are best suited for quantum state transport.

\end{abstract}

\maketitle

%\begin{multicols}{2}
%%% I can see this could take a long time to write a couple of introductory 
%%  paragraphs!

Quantum entanglement has been recognized as an important resource for quantum
information and computation\cite{Nielsen}, for transmission of quantum states
through a channel\cite{Bennet}. 
There have been many proposals of physical systems to
serve as channels for quantum communication\cite{Kielpinski}, and in
particular the spin chains\cite{Bose}.
The essential idea is to encode one particular qubit (spin-1/2
degree of freedom), and let it be transported across the chain to recover the
code from another qubit some distance away. The quantum spin chains are well
suited for state transport, as one could use the Schroeding dynamics
to propagate entanglement, and thus achieve the desired quantum 
communication.   

The dynamics of entanglement of spin systems, viz. the study of time evolution 
of an 
initial state, can be classified into three categories. The dynamics of a $S^z$ 
definite state or a $S^z$ non-definite state under $S^z$ conserving time 
evolution (for example the Heisenberg-XY model), 
and the time evolution of an initial state under a $S^z$ nonconserving dynamics
(for example the transverse Ising model). The structure
of entanglement sharing, the time scales for entanglement transport are very
different between these categories. In this letter, we will study the
Heisenberg-XY model, where $S^z$ is conserved through the time evolution. 
The main features of the dynamics of entanglement 
that will be addressed are, the appearance of pairwise entanglements between
distant spins,
viz. concurrences, starting from an initial state with no entangled pairs or
exactly one entangled pair of spins, and the time scales for the transport of 
entanglement. The maximally
entangled initial states $\uparrow\downarrow \pm \downarrow\uparrow$ (the Bell
state B1) and
$\uparrow\uparrow \pm \downarrow\downarrow$ (the Bell state B2) evolve quite 
differently as we shall see below. The dynamics of entanglement in these states
involve one-magnon and two-magnon excitations. The entanglement sharing in 
one-magnon and two-magnon eignstates of the Heisenberg-XY model has been 
studied\cite{Arul,Subrah}.  The time evolution of B1 states involves the
one-magnon excitations, which are not affected by $s^z-s^z$ interactions, where
as for the B2 states two-magnon excitations are involved, which introduce many 
complications. 
Here, not only two-magnon
scattering states (equivalent to two one-magnon excitations) that have a weak 
signature of the interactions, but two-magnon bound states with a strong
interaction effects\cite{Subrah} have a significant contribution for
entanglement dynamics. 

Let us consider an anisotropic Heisenberg-XY model for a linear chain of 
spins (s=1/2), with
a Hamiltonian
\begin{equation}
H=K_z\sum s_i^zs_{i+1}^z -{K\over2}\sum s_i^+s_{i+1}^- +H.c -B \sum s_i^z -E_0
\end{equation}
where $K_z, K$ are the interaction strengths for the z-components and the
xy components respectively, and $B$ is the strength of the magnetic field along
the z direction. The constant $E_0=-NK_z/4 -NB/2$ is the energy of the
ferromagnetic state, with all the spins poplarized along the z direction. The
above Hamiltonian generates unitary evolution from the Schroedinger equation
which conserves the total $S^z$ in the state. The energy eigenstates and
eigenvalues are known exactly through the Bethe ansatz\cite{Skriyabin}.
The ferromagnetic state $|F\rangle =|\uparrow..\uparrow\rangle$ has no dynamics, being
an eigenstate with zero eigenvalue, and has no entanglement between any pair
of spins. Let us first consider an unentangled state with one spin, 
at the site $l$, in an unknown state, at time t=0 given by
\begin{equation}
|\psi_u(0)\rangle=|\uparrow..\uparrow\rangle|\alpha \uparrow +\beta 
\downarrow\rangle_l\equiv (
\alpha+\beta s_l^-)|F\rangle.
\end{equation} This state has no entanglement, being a direct product state
of different site states.  Let us denote $|n\rangle=s_n^-|F\rangle$, a state with one
down spin at site $n$. At a later time $t$, the state can be written as
\begin{equation}
|\psi_u(t)\rangle=\alpha |F\rangle + \beta \sum \phi_n(t) |n\rangle
\end{equation} where $\phi_n=1/N\sum_q \exp{i(q(n-l)-E_qt/\hbar)}$, and the
one-magnon energy $E_q=-K\cos q + B$. In the limit of large number of spins, $N
\rightarrow \infty$, the wave function can be expressed in terms of the Bessel 
function $J_{n-l}$ as 
\begin{equation}
\phi_n(t)=e^{-i{Bt\over\hbar}}e^{i{\pi\over 2}(n-l)}J_{n-l} ({Kt/\hbar}).
\end{equation} The time scale for the structure in the wave function is
$\tau=\hbar/K$, and from now on we will write the time as a multiple of $\tau$,
as $T=t/\tau$. We can estimate the time scale, using $K\sim 0.01 eV$, as
$\tau\sim 10^{-13}$ sec. 
And the magnetic field adds
on a constant phase to the wave function, and thus can be dropped.
At time $t$ the  mixed state of a given site $i$ can be 
denoted by
the reduced density matrix $\rho_i=tr^{\prime}_{1..N} |\psi_u\rangle\langle
\psi_u|$, where
the prime indicates a partial trace over all states except at site $i$. It is
straightforward to write down the reduced density matrix as
\begin{equation}
\rho_i=(1-|\beta|^2|\phi_i(t)|^2)|\uparrow\rangle\langle\uparrow|+|\beta|^2|\phi_i(t)|^2
|\downarrow\rangle\langle\downarrow|.\end{equation}
Now the initial unknown state encoded in th l'th qubit, can be extracted from
the i'th qubit with a fidelity $F_i= Tr \rho_u\rho_i$ where
$\rho_u=|\alpha \uparrow +\beta \downarrow\rangle\langle\alpha\uparrow+\beta\downarrow|$,
which works out to be (with $i=l+r$, a distance $r$ away from the initial site)
$
F_r(T)=|\alpha|^2 +|\beta|^2(|\beta|^2-|\alpha|^2)J_{r}^2(T).
$ By averaging over all possible initial states, i.e. the Bloch sphere, the
average fidelity is
\begin{equation}
F_r(T)={1\over2}+{1\over6}J_r^2.
\end{equation}
A similar formula has been derived in \cite{Bose}, except our calculation is
simplified due to the limit of large $N$. Now, we are interested in
propagating the initial code to a distance $r$, and recover it. The fidelity
$F_{r}$ has a maximum value for $T\approx r$, which means the quantum state
is transported at a rate $v_t=1/\tau$. After waiting for a time $t=v_tr$ we
have the best recovery of the quantum state at site a distance $r$ away from
the initial site $l$. In Fig.1, the fidelity has been plotted as a function
of time, for $r=100$; the first maximum is for $T_c\approx r$, along with
the result for entangled initial states we shall discuss below.

Though the initial state has no pairwise entanglement, for $T\ne 0$, the
state develops entanglement. We use the concurrence measure\cite{Wootters}
for the pairwise entanglement, which can be calculated from the two-site
reduced density matrix $\rho_{ij}$ (which is obtained by tracing over all
spins except those at sites $i$ and $j$). The time-reversed density matrix
is denoted by $\tilde\rho_{ij}$, and the eigenvalues of $\rho_{ij}\tilde\rho_{
ij}$ by $\lambda_1..\lambda_4$ in the descending order. Then the concurrence
between the two sites is\cite{Wootters},
$C_{ij}={\rm max}(\lambda_1^{1\over2}
-\lambda_2^{ 1\over2}
-\lambda_3^{ 1\over2}
-\lambda_4^{ 1\over2},0)$. Here, the two-site density matrix has the form
$$\rho_{ij}=\left(\begin{array}{cccc}
        1-|\beta|^2(|\phi_i|^2+|\phi_j|^2)& & & \\
         & |\beta|^2|\phi_j|^2&|\beta|^2\phi_i\phi_j^{\star}& \\
         & |\beta|^2\phi_i^{\star}\phi_j&|\beta|^2|\phi_i|^2& \\
        &&&0 \end{array} \right).$$ Now the concurrence is given by\cite{Arul}
$C_{ij}=2|\beta|^2|J_{i-l}(T)J_{j-l}(T)|$. A plot of concurrences vs $T$
are shown in Fig.2, for $i=l+1,j=l$ and $i=l+2,j=l$.
For small $T$, the concurrences grow as $C_{ij}\approx 
= (2|\beta|^2/ r!)(T/2)^r$. 
\begin{figure}     %%%%%%% Figure 1 %%%%%%%%%%%%%%
\input{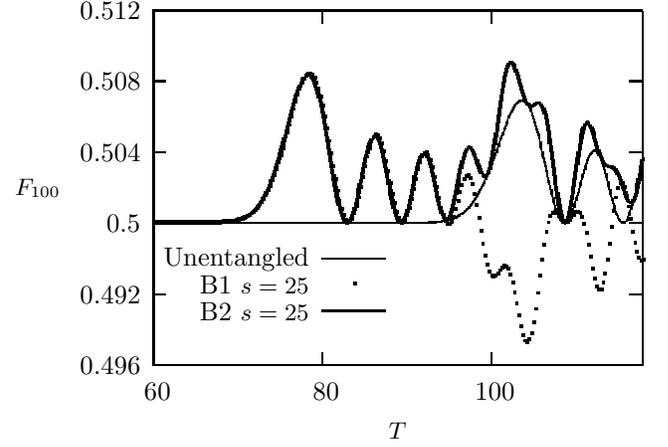}
\caption{The average fidelity $F_r$ as a function of $T$, for $r=100$, for 
the three
types of states discussed. For the entangled states, $s=l-m=25$ has been
used. The first maximum for the unentangled is at $T=r=100$, and for the
B1 and B2 states, at $T=r-s$. }
\end{figure}
For large $T$, we have
$ C_{ij}\approx {2|\beta|^2\over \pi T}$ with oscillations. 

Now we will turn to initial states with entangled pairs of spins. Entangled 
states are expected to better than the unentangled states considered above.
The dynamics will transport and further generate entanglement, as we shall 
see below.
Let us
first consider an initial state with a pair of qubits at sites $l$ and $m$
in an entangled state (B1 state) $\alpha |\uparrow\downarrow\rangle+\beta
|\downarrow\uparrow\rangle$, all other spins polarized, which is
represented as
\begin{equation}
|\psi_1(T=0)\rangle=(\alpha s_l^-+\beta s_m^-)|F\rangle=\sum \phi_n(0)|n\rangle.\end{equation}
For $T=0$
all concurrences are zero except, $C_{lm}=1$. For $T\ne 0$, this 
entanglement spreads, and is transported to other pairs. Again using the
one-magnon excited states, we can write down the wave function as a function
of $T$  
\begin{equation}
\phi_n(T)=\beta e^{i{\pi\over2}(n-l)}J_{i-l}(T)+\alpha e^{i{\pi\over2}(n-m)} 
J_{i-m}(T).
\end{equation}
Since the above state is a one-magnon state (though may not be an eigenstate)
the concurrence is given by\cite{Arul}
$ C_{ij}=2|\phi_i^{\star}\phi_j|$. In particular, the concurrence between
the sites $l$ and $m$ for later times, for the maximally-entangled initial
state ($\alpha=\beta=1/\sqrt{2}$), for $l-m$ even or odd respectively, is
given as
\begin{equation}C_{lm}=(J_0+(-1)^{l-m\over2}J_{l-m})^2~~{\rm or~~} (J_0^2+J_{l-m}
^2).\end{equation}
In Figure 2, $C_{lm}$ has been plotted as a function of $T$, for $l-m=1$,
which drops from the initial value of unity as $T^{-2}$. 

Now, the fidelity of recovering a quantum state $|\alpha\uparrow+\beta
\downarrow\rangle$ at a site $i$ can be calculated straightforwardly, analogous to
the unentangled state we calculated before, as 
(for $i=m+r$, a distance $r$
away from the site $m$)
\begin{figure}   %%%%%%%% Figure 2 %%%%%%%%%%%%%%
\input{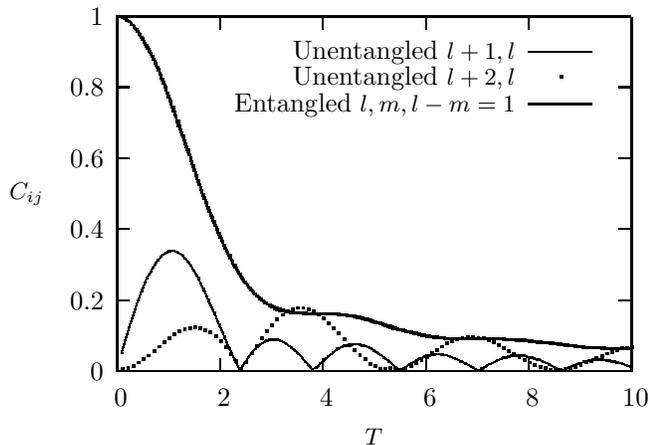}
\caption{The concurrence $C_{ij}$ as a function of $T$, using $|\beta|^2=1/2$.
It can be seen how  concurrence builds up between the initial site $l$, 
$l+1$ and $l+r$ for the unentangled initial state, and how the concurrence
decreases between the initial sites $l$ and $m$ for B1 and B2 Bell states. }
\end{figure}
$ F_r(T)=|\alpha|^2+(|\beta|^2-|\alpha|^2)|\phi_{r+m}(T)|^2.$ For $|\alpha|
\approx |\beta|$, the fidelity is close to 1/4, as before for the unentangled
case, which means the channel is noisy for a good recovery. In this case the
unknown state can be recovered from two sites, as we shall discuss below.
The average fidelity, after averaging over
the Bloch sphere, is given by 
\begin{equation}
F_r(T)={1\over2}+{1\over6}( J_{r-(l-m)}^2
-J_r^2)
\end{equation}
which should be compared with the formula we obtained (Eq.6) for the 
unentangled initial state. The first maximum of the fidelity now depends on
$J_{r-(l-m)}$, which will occur for an earlier time, and the maximum value will
be more than for $J_r$. The rate at which the state propagates is still
$v_t=1/\tau$, but we need to wait for a shorter time interval $t=v_t(r-(l-m))$ 
to recover the state
a distance $r$ away from the $m'$th site. The distance between the two initial
sites can be chosen conveniently for a given distance over which the transport
is desired. The fidelity for this case has been
shown in Figure 1 for a few values of $r$ as a function of time. Due to the
presence of the competing term (the last term in the above), the fidelity
could fall below the value for the unentangled state, after the first maximum. 

The initial entangled state at sites $l$ and $m$,$\rho_{B1}=|\alpha\uparrow
\downarrow+\beta\downarrow\uparrow\rangle\langle\alpha\uparrow\downarrow+
\beta\downarrow \uparrow\rangle$,
can be extracted from sites $i$ and $j$ at a later time with a fidelity, 
$G_{ij}=Tr \rho_{B1}\rho_{ij}$, which
can be calculated after some manipulations simply as 
$
G_{ij}=|\alpha\phi_j+\beta\phi_i|^2.$ 
This has a maxima structure even for
$|\alpha|\approx|\beta|$, unlike the fidelity function $F_i$ discussed
earlier.
For $i=l+r,j=m+r$, i.e. for
a pair of sites translated by $r$ from the initial pair $l,m$, the fidelity
takes the form (using $s=l-m$), 
$G_r=|2\alpha\beta J_r +\alpha^2\exp{(i\pi s/2)}J_{r+s}+\beta^2
\exp{(-i\pi s/2)}J_{r-s}|^2$. 
After averaging over the Bloch sphere, 
\begin{figure}  %%%%%%%%%% Figure 3 %%%%%%%%
\input{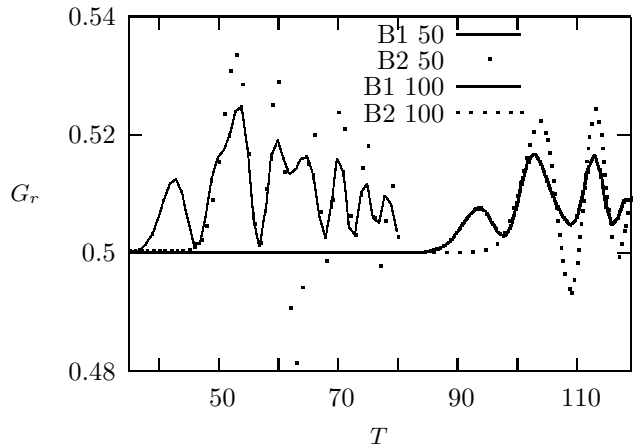}
\caption{The average fidelity $G_r$ as a function of $T$, for $r=50,100$ using
$s=l-m=10$. A constant 0.5 has been added on for B1 state, to show both on
the same graph. }
\end{figure}
it has a simple form  
\begin{equation}
G_{r} ={1\over3}(J_{r-s}^2+J_{r+s}^2)+{2\over3}J_r^2
\end{equation}
In Fig.3, we show the fidelity $G_r$ as a function of $T$ for 
$r=50,100$.

Let us now turn to the most difficult case of an initial state with a
B2 state at sites $l$ and $m$, and the rest of the spins polarized,
given as $(\alpha+\beta s_l^-s_m^-)|F\rangle\equiv \alpha|F\rangle+\beta|\Phi(T=0)\rangle$. 
The state can be written as
\begin{equation}
|\psi_2(T=0)\rangle=
\alpha|F\rangle+\beta\sum \phi_{ij}(0)|ij\rangle,
\end{equation}
where $|ij\rangle$ stands for a two-magnon basis state with two down spins at sites 
$i$ and $j$. Initially for $i=l,j=m$ the wave function is unity, and zero for
all other values of $i$ and $j$.
The time evolution of the second term above can be worked out in terms of the 
two-magnon excitations, using the Bethe Ansatz solution for two down spins.
The concurrences in the two-magnon eigenstates have been worked out\cite{
Subrah} both for the scattering and the bound states.
The wave function is extremely complicated due to the magnon interactions
arising for a nonzero $K_z$,
which can be vastly simplified by taking the limit $K\rangle\rangle K_z$, i.e. dropping
the interaction $K_z$ terms in the Hamiltonian. In the XY limit, the
wave function as a function of time takes the form (after taking $N\rightarrow
\infty$ limit)
\begin{equation}
\phi_{ij}(T)=e^{i{\pi\over2}(i+j-l-m)}( J_{i-l}(T)J_{j-m}-J_{i-m}J_{j-l}),
\end{equation}
which is antisymmetric in the two indices, reflecting the underlying fermionic
nature of the moving down spins. 
Now,
following through the steps as before, the two-site reduced density matrix
has the form
$$\rho_{ij}=\left(\begin{array}{cccc}
        |\alpha|^2+\beta|^2u_{ij}&&&\alpha\beta^{\star}\phi_{ij}^{\star}\\
        & |\beta|^2w_{1ij}&|\beta|^2z_{ij}^{*}& \\
         &|\beta|^2z_{ij}&|\beta|^2w_{2ij}& \\
        \alpha^{\star}\beta\phi_{ij}& & &|\beta|^2v_{ij} \end{array} \right).
$$
In the above, the various matrix elements stand for
$u_{ij}=\langle ({1/2}+s_i^z)({1/2}+s_j^z)\rangle,
v_{ij}=\langle ({1/2}-s_i^z)({1/2}-s_j^z)\rangle,
w_{1ij}=\langle ({1/2}-s_i^z)({1/2}+s_j^z)\rangle,
w_{2ij}=\langle ({1/2}+s_i^z)({1/2}-s_j^z)\rangle,
z_{ij}=\langle s_j^+s_i^-\rangle,$
where the expectation value is taken in the two-magnon state $|\Phi(T)\rangle$ only.
Now following through the further steps of Wootters\cite{Wootters}, the
eigenvalues of $\rho_{ij}\tilde\rho_{ij}$ for the above density matrix are
$(\sqrt{(|\alpha|^2+u|\beta|^2|)\beta|^2v}\pm |\alpha^{\star}\beta\phi_{ij}
|)^2,|\beta|^4(\sqrt{w_1w_2}\pm |z|)^2$. This gives two regimes for the
concurrence as
\begin{eqnarray}
C_{ij}&&=2|\beta|^2|z_{ij}|-2|\beta|\sqrt{v_{ij}}\sqrt{|\alpha|^2+|\beta|^2u_{
ij}} ~~{\rm or}  \nonumber\\
&&=2|\alpha^{\star}\beta\phi_{ij}|-2|\beta|^2\sqrt{w_{1ij}w_{2ij}}
\end{eqnarray}
which ever term is positive, and otherwise zero. The off-diagonal matrix 
element can be calculated as (taking $i\rangle j$)
\begin{equation}
z_{ij}=\sum 
\phi_{in}^{\star}\phi_{jn} 
- 2 \sum_{j+1}^{i-1} \phi_{in}^{\star}\phi_{jn}\equiv \eta_{ij}-2\zeta_{ij}
\end{equation}
where the sum in $\eta_{ij}$ is over all values of $n$,  which can be 
calculated using the addition rule $J_n(x+y)=\sum_k J_k(x)J_{n-k}(y)$, and
settig the site $m$ at the middle of the chain for convenience, 
\begin{equation}
\eta_{ij}=e^{i{\pi\over2}(j-i)}(J_{i-l}J_{j-l}+J_{i-m}J_{j-m}).
\end{equation}
And $\zeta_{ij}$, which is just the finite sum, is
quite complicated to calculate in general. The
diagonal matrix elements
are $v_{ij}=|\phi_{ij}|^2,u_{ij}\approx 1, w_{1ij}=\eta_{ii} -
|\phi_{ij}|^2, w_{2ij}=\eta_{jj}-|\phi_{ij}|^2.$ This simplifies the
expression for the concurrence between two sites $i$ and $j$ as
\begin{equation}
C_{ij}=|\beta|{\rm max}(0,|\beta||z|-|\phi|,2(|\alpha\phi|-
|\beta|\sqrt{w_{1}w_{2}})).\end{equation}
For $l-m$ odd, the off-diagonal matrix element $z_{lm}=0$, and the expression 
for the maximally-entangled initial state is simple,
$
C_{lm}=J_0^2+J_{l-m}^2$, which is what we got for B1 Bell states. For $l-m$
even, the expression is still complicated. For $l-m=2$, $C_{lm}={\rm max}(0,|z|
/2-|\phi|/\sqrt{2}, |\phi|-w_1)$, where
$\phi=(J_0^2-J_2^2),w_1=J_0^2+J_2^2-|\phi|^2,z=2J_0J_2+J_1^2(J_0+J_2)^2$. 
The concurrence between the sites $l$ and $m$ is plotted as a function of
$T$, for $l-m=1$ in Fig.2, along with the result for the B1 states and the
unentangled state. 

The fidelity of recovering the state $|\alpha\uparrow+\beta\downarrow\rangle$ at
a site $i$ is again straightforward, $F_i=|\alpha|^2+|\beta|^2 (|\beta|^2-
|\alpha|^2)\eta_{ii}$. The average fidelity, for $i=m+r$,
\begin{equation}
F_r={1\over2}+{1\over6}(J_{r-(l-m)}^2+J_r^2).
\end{equation}
As compared to the expression for the B1 states, in the above there is no
competition between $J_r$ and $J_{r-(l-m)}$. The fidelity here is
greater than that of the unentangled state for all times. The first maximum
value is determined by the first term, as in the case of B1 states, which
occurs for $t=v_t (r-l+m)$. A comparison of the fidelity
as a function of $T$ for all the three cases, from Fig.1 for $r=100$,
shows that the B2 states have better fidelity.  
Now, the fidelity of recovering $|\alpha\uparrow\uparrow+\beta\downarrow
\downarrow\rangle $ from sites $i$ and $j$, analogous to the B1 states
we discussed before, is given as
$G_{ij}=|\alpha|^2+|\beta|^4|\phi_{ij}|^2+|\alpha\beta|^2(\phi_{ij}^{\star}+
\phi_{ij})$. This function also exhibits a maxima structure for 
$|\alpha|\approx |\beta|$. The average fidelity is (for $i=l+r,j=m+r$)
\begin{equation}
G_r={1\over2}+{1\over3}(J_r^2-J_{r-s}J_{r+s})(
J_r^2-J_{r-s}J_{r+s} +e^{i\pi r}).
\end{equation}
The fidelity as a function of $T$ is shown in Fig.3, for $r=50,100,s=10$,
along with the result for B1 states. The B2 states exhibit better fidelity here
also, as is the case in Fig.1.

In conclusion, we have investigated the quantum state transport across a
channel of qubits, the spin chain, using the Heisenberg-XY dynamics. The
presence of entanglement and its dynamics is crucial for communication
over the channel. Initial states with a pair of qubits in a state
$\alpha\uparrow\uparrow+\beta\downarrow\downarrow$ show better fidelity.
Here, it will be interesting to investigate the effect of a nonzero $K_z$;  
significant changes in the wave functions, the concurrences and the fidelity
are expected. Finally,   
states with many entangled pairs in an optimized network may 
demonstrate an almost-ideal quantum communication, that is, a teleportation 
protocol with the sender and the receiver at a fixed distance and the
code transported with negligible interference from the network channel.

\end{document}